\documentclass[%
notitlepage,
reprint,
superscriptaddress,
amsmath,amssymb,
aps,
]{revtex4-2}

\usepackage{lipsum}
\usepackage{graphicx}
\usepackage{dcolumn}
\usepackage{bm}
\usepackage{hyperref}

\usepackage{epstopdf}

\newcommand{\bra}[1]{\langle #1 \vert}
\newcommand{\ket}[1]{\vert #1 \rangle}

\begin{document}
	\title{Rogue waves in quantum lattices with correlated disorder}

\author{A. R. C. Buarque}
\affiliation{%
	Instituto de F\'{i}sica, Universidade Federal de Alagoas, 57072-900 Macei\'{o}, AL, Brazil
}%
\affiliation{%
	Secretaria de Estado da Educação de Alagoas, 57055-055, Macei\'{o}, AL, Brazil
}%
\author{W. S. Dias}
\affiliation{%
	Instituto de F\'{i}sica, Universidade Federal de Alagoas, 57072-900 Macei\'{o}, AL, Brazil
}%
\author{G. M. A. Almeida}
\affiliation{%
	Instituto de F\'{i}sica, Universidade Federal de Alagoas, 57072-900 Macei\'{o}, AL, Brazil
}%
\author{M. L. Lyra}
\affiliation{%
	Instituto de F\'{i}sica, Universidade Federal de Alagoas, 57072-900 Macei\'{o}, AL, Brazil
}%
\author{F. A. B. F. de Moura}
\email{fidelis@fis.ufal.br}
\affiliation{%
	Instituto de F\'{i}sica, Universidade Federal de Alagoas, 57072-900 Macei\'{o}, AL, Brazil
}%

\begin{abstract}

We investigate the outbreak of anomalous quantum wavefunction amplitudes in a one-dimensional 
tight-binding lattice featuring correlated diagonal disorder. Such rogue-wave-like behavior is fostered by
a competition between localization and mobility. The effective correlation length of the disorder is ultimately
responsible for bringing the local disorder strength to a minimum, fuelling the occurrence
of extreme events of much higher amplitudes,  
specially when compared to the case of uncorrelated disorder. 
Our findings are valid for a class of discrete one-dimensional systems and reveal profound
aspects of the role of randomness in rogue-wave generation. 
\end{abstract}
\maketitle
	
\section{Introduction} 

Rare and unpredictable events carrying huge impact are widespread in nature, from stock markets to physical sciences.
Outliers may change the course of things 
more often than we are prone to think, sometimes leading to hazardous consequences. 
One example is the emergence of rogue waves in the ocean. 
The famous Draupner wave recorded in 1995 at a gas platform in Norway was twice as big as the significant wave height of the area. This
happened to be the first scientific
evidence of a rogue wave \cite{harvey04} and
led to a burst of 
interest in the field as studies began to suggest that these extreme events would occur more frequently
than assumed from ordinary Gaussian statistics \cite{dudley19}. 
About a decade later, Solli \textit{et al.}  introduced rogue waves in optics based on observations made on fibre 
supercontinuum generation in the presence of noise \cite{solli07}.

The analogy drawn between oceanic rogue waves 
and extreme instabilities in optics associated to long-tailed statistics 
set the stage for a number of theories aimed to explain the physical mechanisms behind rogue-wave generation.
Much of the effort has been directed toward establishing whether and which linear or nonlinear processes play
the biggest roles \cite{bonatto11, baronio14,xu20, akhmediev09, onorato01, ying11,safari17, arecchi11}. Oceanic rogue waves, for instance, may result from
various mechanisms in action, such as constructive interference of random fields,
modulational instability, and soliton modes,
depending on sea and wind conditions \cite{kharif_book, dysthe08}. 
Even though there is no definitive consensus on that matter, neither robust ways to predict
where and when rogue waves will occur, 
noise and randomness seem to be key ingredients for their occurrence \cite{hohmann10, arecchi11, liu15} (deterministic rogue waves are also discussed in Refs. \cite{bonatto11, baronio14}).

Some degree of disorder is paramount for generating rogue waves via linear mechanisms in particular \cite{hohmann10, arecchi11, metzger14, mathis15, liu15, mattheakis16, derevyanko18, peysokhan19, rivas20, bonatto20, frostig20, dasilva20, buarque22}. 
Very recently, this was addressed in the context of quantum walks \cite{buarque22}.
Therein, the authors primarily sought to explore the (hitherto elusive) relationship between Anderson localization and rogue wave manifestation.
They reported a minimal disorder threshold $\propto N^{-1/2}$ ($N$ being the number of sites) above which rogue waves were created, whereas
intermediate disorder levels would maximize the chances of seeing one due to a proper balance between trapping and mobility (see also Ref. \cite{rivas20}). 
Such results, besides bringing the rich subject of rogue waves up to the realm of quantum transport, add important elements 
to the issue of the actual role played by randomness. 

With those ideas in mind, here we set out to track the dynamics of rogue waves
on a single quantum particle propagating in a lattice featuring correlated disorder. Anderson localization theory
settles that all single-particle eigenstates are exponentially localized for any amount of uncorrelated disorder in one and two dimensions \cite{abrahams79}. This can be violated, however, when the disorder displays intrinsic correlations \cite{demoura98, dietz11,izrailev_rev}.
Scale-free correlations, for instance, are known to support a metal-insulator transition with sharp
mobility edges \cite{demoura98}.

In this article, our goal is to investigate the 
development of sudden, anomalous quantum amplitudes 
due to the interplay between localized and extended states.
Indeed, long-range correlated fluctuations
in the random input phases was recently 
shown
to produce
rogue waves way above the threshold in an experiment on linear
light diffraction in 1D \cite{bonatto20}. 

We look at a particular kind of correlated disorder in which 
a single parameter is able to control the typical correlation length and, in turn, the local disorder strength.
The latter is found to be a crucial factor underlying the generation of the rogue waves for it sets up the
right amount of wavefunction mobility. This boosts not only the number of occurrences, but also
the average rogue wave amplitude.

\section{Model}

We consider a single quantum particle propagating in a one-dimensional array described by the tight-binding Hamiltonian
\begin{eqnarray}\label{equation1}
H=J\sum_{n=1}^{N}(\ket{n}\bra{n+1} + \ket{n+1}\bra{n}) +
\sum_{n}^{N}\varepsilon_{n}\ket{n}\bra{n},
\end{eqnarray}
where $J$ is the nearest-neighbor hopping strength, $\varepsilon_{n}$ is the on-site potential, and states $\ket{n}$ represent the location of the particle and span the whole Hilbert space. 
As such, an arbitrary quantum state can be written as
$\ket{\Psi}=\sum_{n}\psi_{n}\ket{n}$, with the normalization condition $\sum_n |\psi_n|^2=1$.  The time evolution of the wave function $\psi_n$ is given by the Schr\"{o}dinger equation 
\begin{eqnarray}\label{equation2}
i\frac{d}{dt}\psi_{n} = \psi_{n+1} + \psi_{n-1} + \varepsilon_{n}\psi_{n},
\end{eqnarray}
where $\hbar=J=1$ without loss of generality. 

Here we go beyond the standard case of uncorrelated disorder and consider that 
the local potentials $\varepsilon_n$ are embedded with correlations, as given by
\begin{eqnarray}\label{equation3}
\varepsilon_{n}=\sum_{m}\frac{Z_{m}}{(1+d_{n,m}/A)^2},
\end{eqnarray}
where $Z_{m}$ is a random number $\in [-1,1]$, $d_{n,m}$ is the Euclidean distance between sites $n$ and $m$, 
and $A$ controls the correlation length of the series. 
We impose periodic boundary conditions, $\ket{N+1} = \ket{1}$, such that 
each site $n$ can be identified through
angular and Cartesian coordinates, $\theta_n=(2\pi/N)n$ and $(x_n=R\sin{\theta_n}, y_n=R\cos{\theta_n})$, respectively,  
rendering $d_{n,m}=\sqrt{(x_n-x_m)^2+(y_n-y_m)^2}$.  The disordered sequence is further normalized to have zero mean and unit variance.

\begin{figure}[t!]
\centering
\includegraphics[scale=0.39,clip]{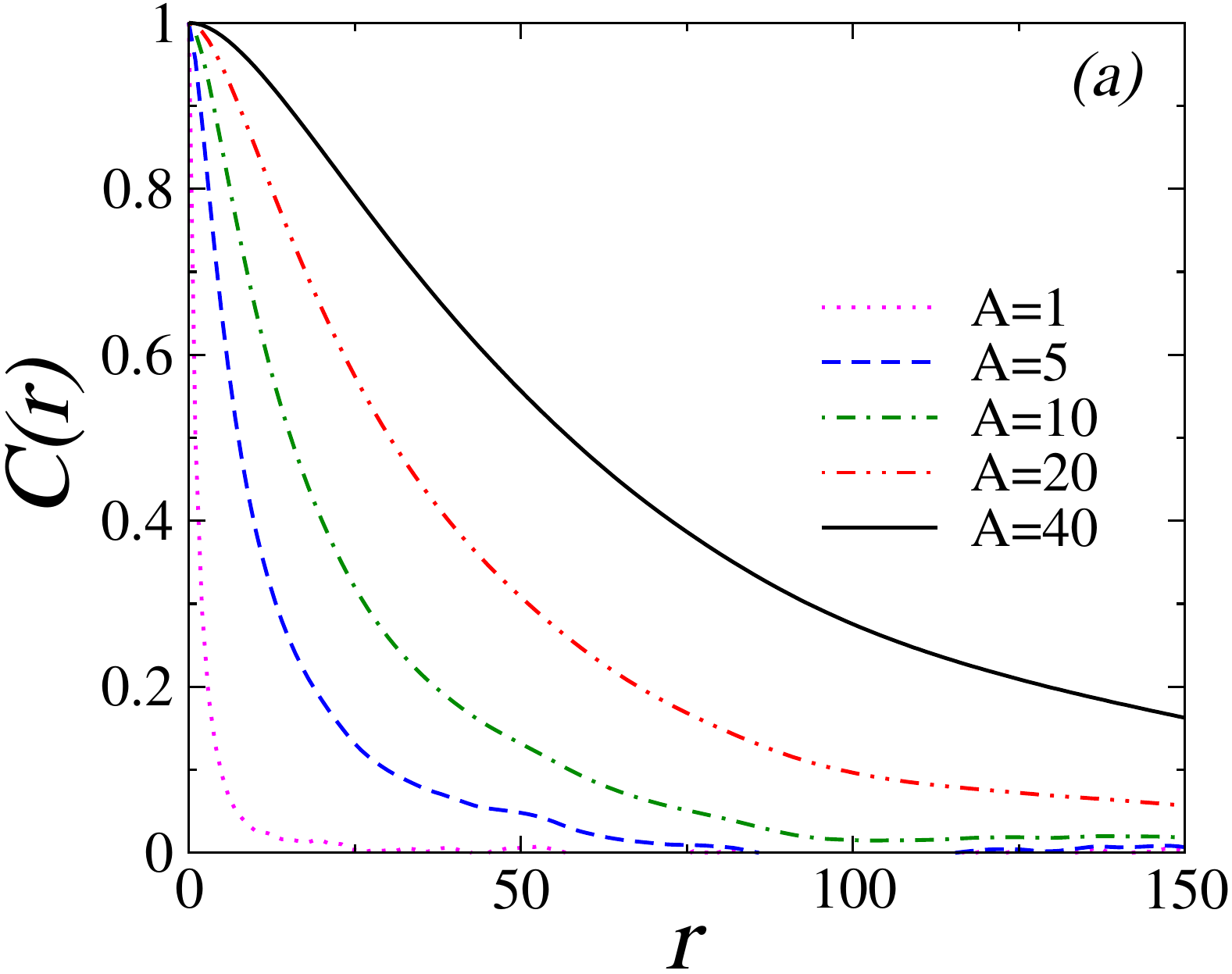}
\hspace*{-0.3cm}\includegraphics[scale=0.38,clip]{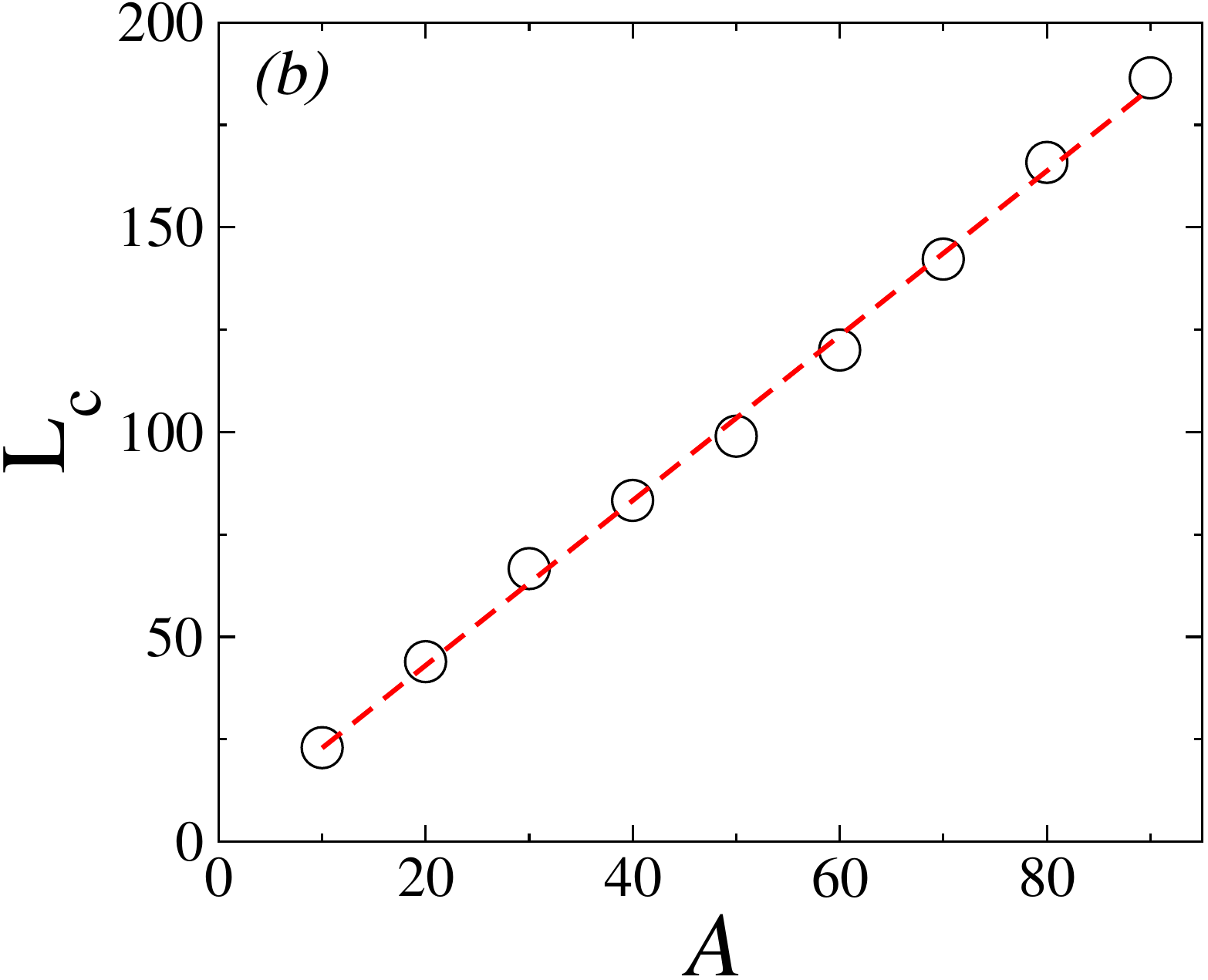}
\caption{(a) Autocorrelation function $C(r)= \mathrm{cov}(\varepsilon_i , \varepsilon_{i+r})$ 
versus $r$, for many values of the correlation parameter $A$, averaged over 100 independent samples.  
(b) Effective correlation length $L_c \propto A$.}
\label{fig1}
\end{figure}

To see how the above disorder configuration play out with the correlation parameter $A$, in Fig. \ref{fig1}(a)
we plot the
autocorrelation function $C(r) =  \mathrm{cov}(\varepsilon_i , \varepsilon_{i+r}) =\sum_{i=1}^{N-r}  \varepsilon_i\varepsilon_{i+r}/(N-r) $.
Upon increasing $A$, $C(r)$ exhibits a slower decay, as expected from Eq. \ref{equation3}. We are then able to set an effective correlation length $L_c$ 
by fitting $C(r)\propto e^{-r/L_c}$. Figure \ref{fig1}(b) shows that $L_c \propto A$ thereby 
affirming what the latter stands for. 

\section{Results}

\begin{figure}[t!]
\centering
    \resizebox{5.2cm}{18.0cm}{\includegraphics{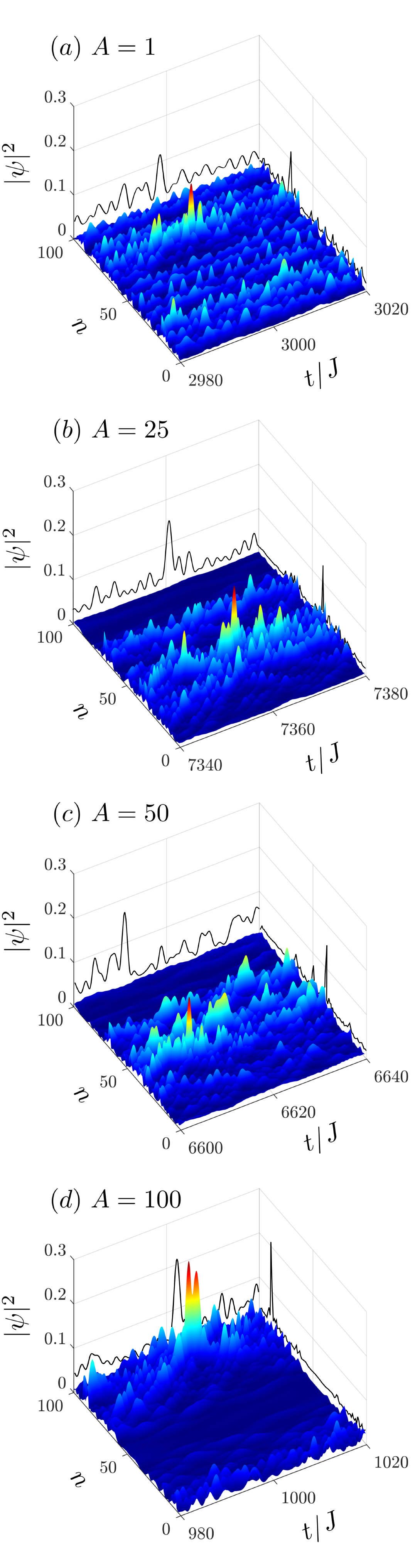}}
    \caption{Snapshots of the space-time evolution of the particle wavefunction probability 
highlighting typical rogue wave events for different correlation strengths (a) $A=1$, (b) $A=25$, (c) $A=50$, (d) $A=100$. 
2D graphs display spatial and temporal cross-sections for the largest amplitudes. Time is expressed in units of $J^{-1}$.
}
    \label{fig2}
\end{figure}  

We are ready to search for the occurrence of rogue waves and address the role of the correlated disorder. 
In all simulations bellow, 
the set of equations given in Eq. \ref{equation2} is numerically solved by 
employing a high-order Taylor expansion of the time evolution operator:
\begin{equation}
U(\Delta t) = \exp (iH\Delta t) = 1 + \sum_{l=1}^{n_{0}}\frac{(iH\Delta t)^l}{l!},
\end{equation}
with time step  $\Delta t = 0.01J^{-1}$ and $n_0 = 20$, which is enough to 
produce smooth outcomes and
keep the norm conserved during the whole time interval. 
In order to avoid ambiguity between a rogue wave event 
and trapping of the wavefunction due to Anderson localization we set 
$\psi_n(t=0) = 1/\sqrt{N}$ for all $n$ \cite{buarque22}.

Figure \ref{fig2} shows typical scenarios of rogue waves, as told by the probability amplitude 
of the particle wavefunction $|\psi_n(t)|^2$, for different values of the correlation strength $A$.
For weak correlations the evolution is characterized
by sparse, low-amplitude waves that eventually add up to produce the rogue events, as shown in Fig. \ref{fig2}(a) 
around $t=3000/J$.
As the correlation strength is increased, the background becomes more inhomogeneous -- a fundamental trait in the generation
of rogue waves via linear processes \cite{arecchi11}.
There sets in distinct amplitude domains in space, indicating that $A$ is pushing for
weaker local fluctuations in the disorder distribution (to be addressed in a moment).
As a consequence, rogue waves of exceptionally higher amplitudes are likely to occur [cf. Fig. \ref{fig2}(d)].

We will not be using here any particular
rogue-wave criteria. One measure employed in various contexts is the significant wave height, commonly defined
as twice the average of largest one third of values in a data set \cite{kharif_book}. An event is thus considered a rogue wave
whenever it beats that level.
Considering our initial state
and the results obtained in Ref. \cite{buarque22}, that deals with a similar class of problem, 
such threshold would be of the order of $1/N$. As our following analysis is built on extreme-value statistics, 
the data is heavily loaded with amplitudes well above that mark. 

\begin{figure}[t!]
    \centering
    \resizebox{8.5cm}{6.cm}{\includegraphics{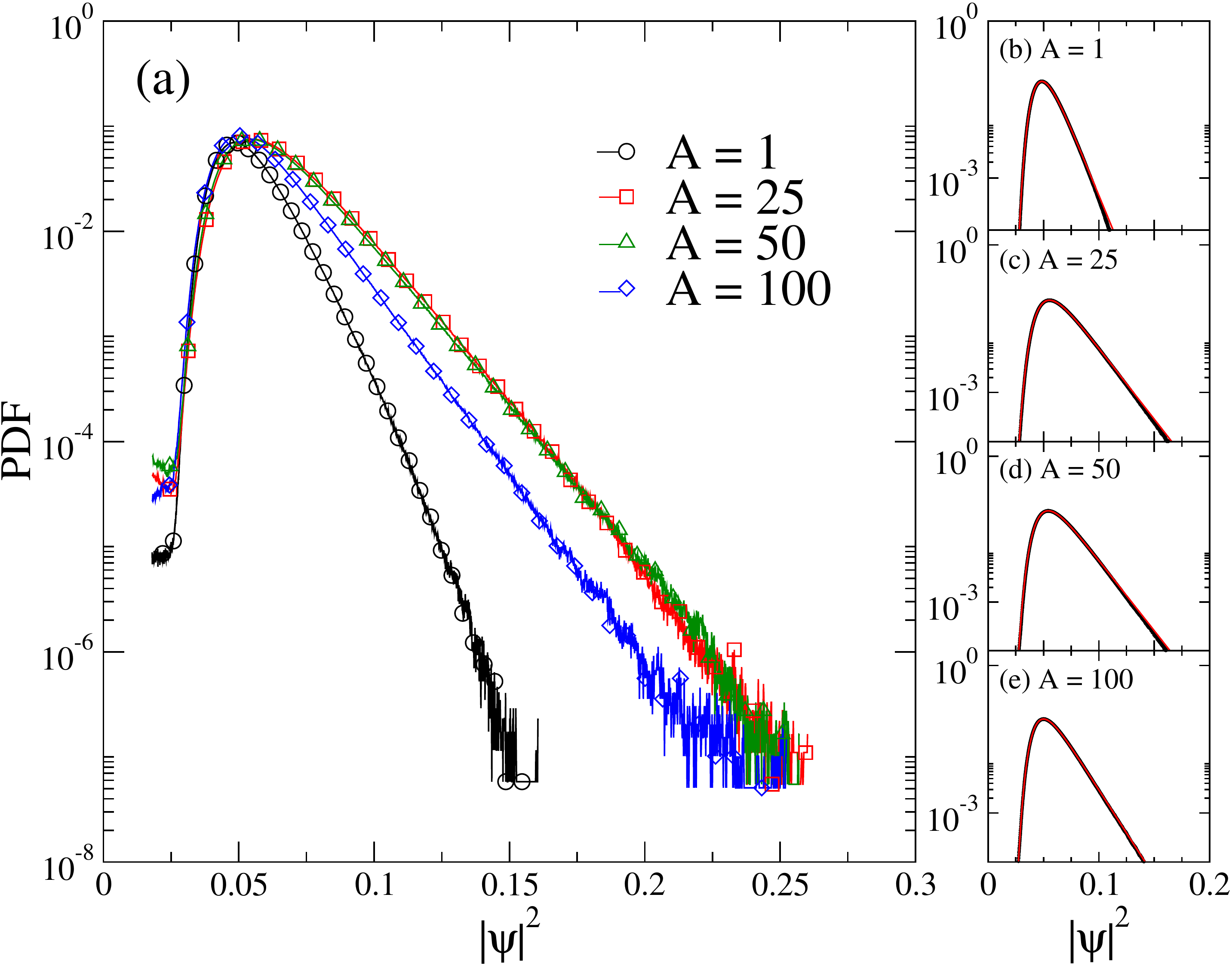}}
    \caption{Extreme-value distributions 
generated by sorting up the maximum values of $|\psi_n|^2$ at each time step during the whole evolution (up to $t=15000/J$ with $\Delta t = 0.01/J$)
for 500 independent samples of a disordered chain with $N=100$ sites when $A=1, 25, 50, 100$. Right panels show 
the fitted Gumbel distribution (red curve) of the form $P(x)\propto \exp[-(\alpha x+\beta \exp(-\alpha x))]$. 
 }
    \label{fig3}
\end{figure}

\begin{figure}[t!]
\begin{center}
\includegraphics*[scale=0.55,clip]{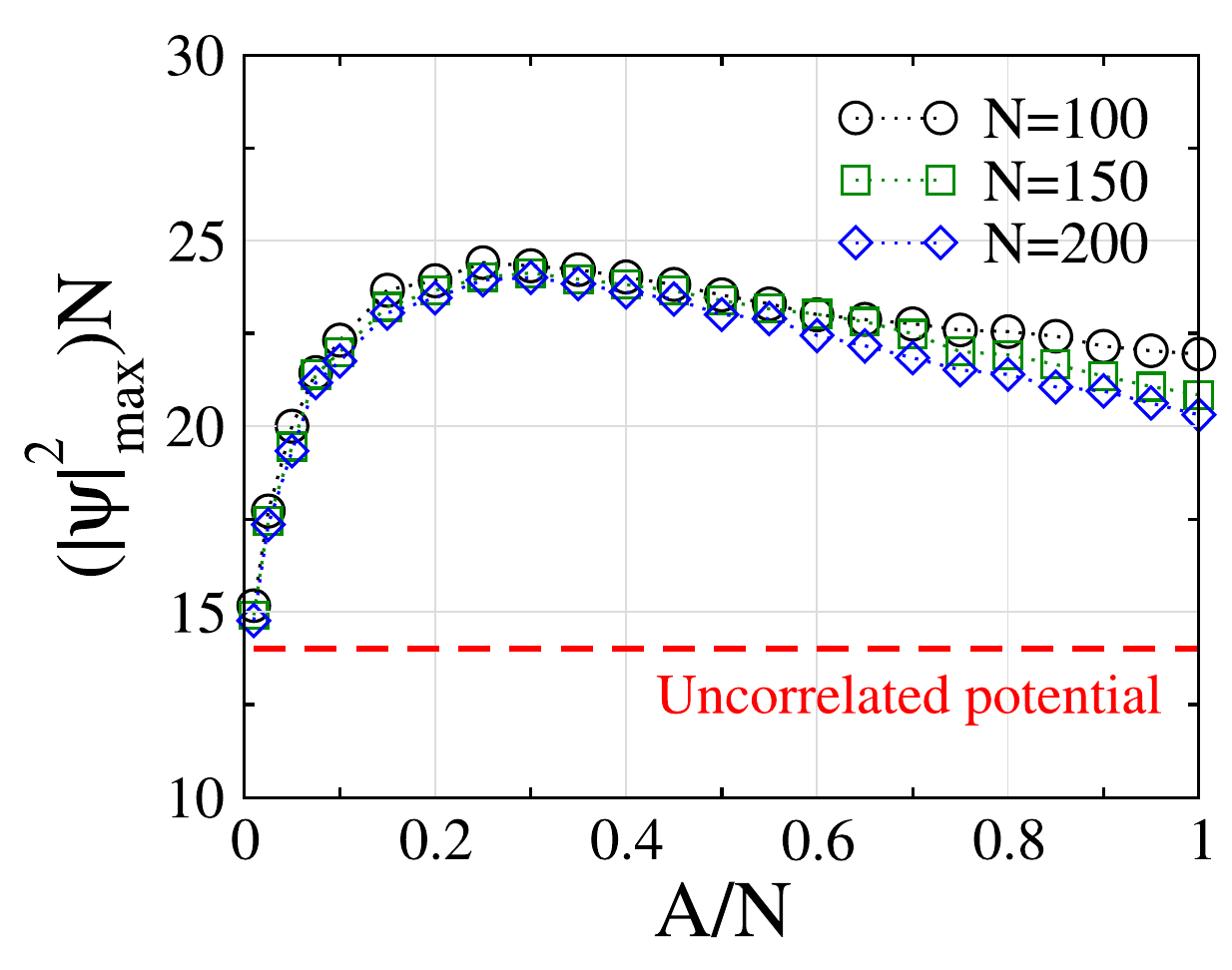}
\end{center}
\caption{Maximum wavefunction probability amplitude $|\psi|_{max}^2$ scaled with $N$ (excluding exceptionally rare events)
for a range of $A/N$ values.}
\label{fig4}
\end{figure}

Let us now obtain one of the limiting extreme-value distributions,
according to the Fischer-Tippett-Gnedenko theorem \cite{majumdar20}.
In order to do so, we pick the maximum wavefunction amplitude in space at each time step for several independent realizations of the disorder.  
The stacked outcomes are seen in the probability density functions (PDFs) shown in Fig. \ref{fig3} for different degrees of correlation $A$. Note that this 
effective correlation length indeed brings forth higher amplitudes. However, there is a threshold value for $A$ above which
the right tail of the distribution begins to deflate (rare, extreme rogue waves can still develop). 
To learn more about this non-monotonic relationship between the rogue-wave maximum amplitudes 
and the correlation parameter $A$, we portray in Fig. \ref{fig4}
the maximum amplitudes $|\psi|_{max}^2$ allowed at a
probability level just above $10^{-7}$ (see Fig. \ref{fig3}). This is done in order to avoid extremely rare outcomes. 
The scaling with $N$ in Fig. \ref{fig4} is employed so 
we can filter out finite-size effects and 
focus on the role of the correlation degree only. 
Compared with what one would obtain from
uncorrelated series of the potential  $\varepsilon_n$ (cf. dashed curve in Fig. \ref{fig4}), the rogue waves 
can reach almost twice as large amplitudes when supported by the correlated disorder. 

We also find that all the skewed PDFs shown in Fig. \ref{fig3}
belong to the Gumbel class of extreme value distributions, fitted by $P(x)\propto \exp[-(\alpha x+\beta \exp(-\alpha x))]$.
This is expected since the tail of the parent function $p(|\psi|^2)$ decays exponentially (slower than that of a Gaussian distribution, 
which is another signature of rogue waves). Similar behavior
is found in quantum walks featuring uncorrelated disorder \cite{buarque22}, meaning that the wavefunction fluctuations are
well described
by processes involving independent and identically distributed random variables.

\begin{figure}[t!]
\begin{center}
\includegraphics[scale=0.45,clip]{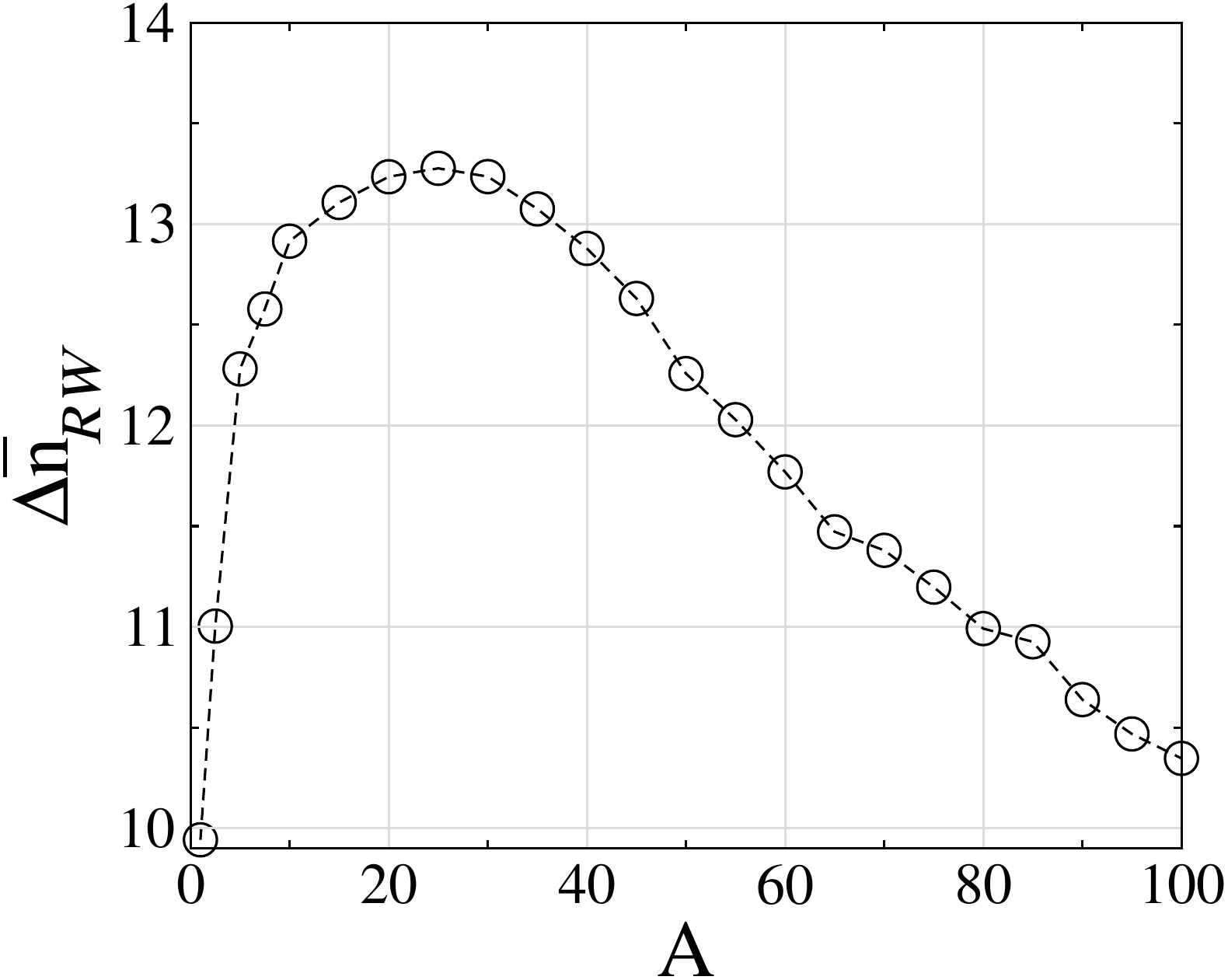}
\end{center}
\caption{Mean number of sites actively involved in the generation of rogue waves $\Delta\overline{n}_{RW}$ versus correlation strength $A$, averaged over 500 independent realizations of disorder, for times up to $t=15000/J$ and $N=100$. Note that the 
values of $A$ that give
highest $\Delta\overline{n}_{RW}$ also leads to outbreak of much intense rogue waves. 
}
\label{fig5}
\end{figure}

\begin{figure}[t!]
\begin{center}
\includegraphics[scale=0.41,clip]{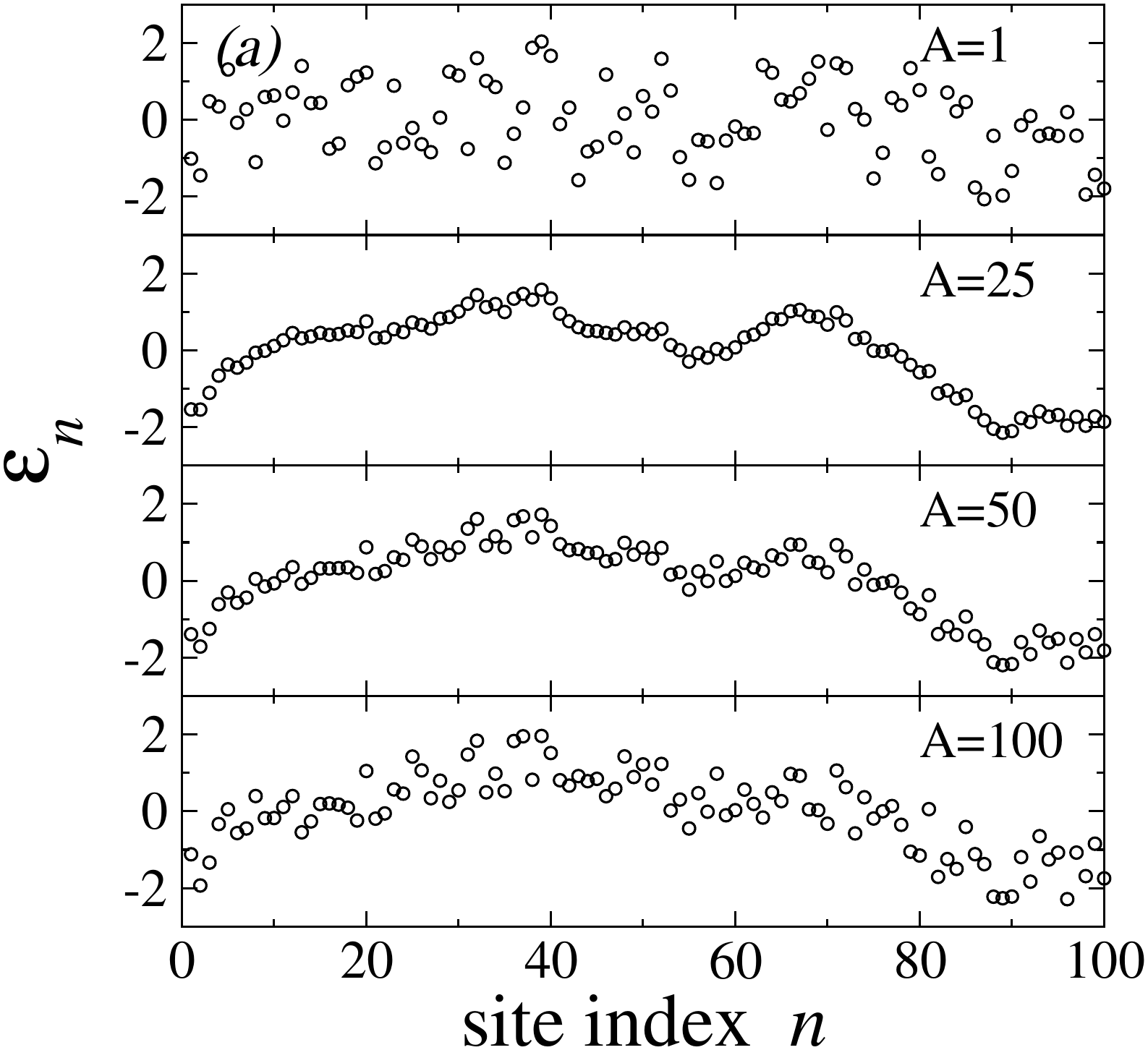}
\includegraphics[scale=0.41,clip]{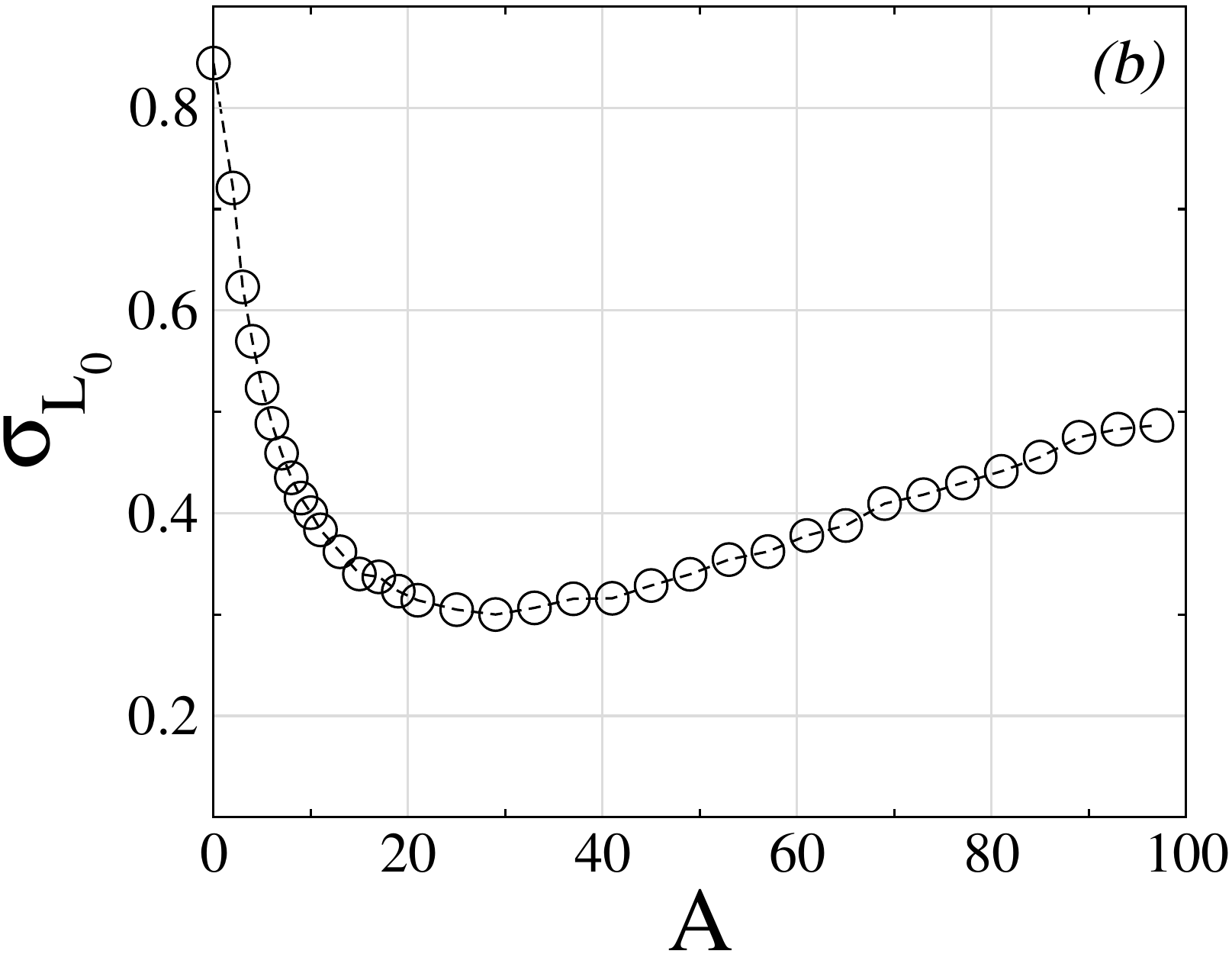}
\end{center}
\caption{(a) Series of the on-site potential  $\varepsilon_n$ considering  $A = 1, 25, 50$, and $100$. (b) Local standard deviation $\sigma_{L_{0}}$ as a measure of the local disorder strength. Weaker local fluctuations
are obtained at intermediate values of $A$, the same range that renders better chances of observing rogue waves
of higher amplitudes.}
\label{fig6}
\end{figure}

The regularity of rogue wave events 
occurring on systems featuring static disorder can 
be predicted to some extent based on the
energy resonance conditions across the lattice. Either very weak or very strong
levels of disorder 
should suppress the onset of anomalous wavefunction fluctuations. Some studies report 
that an optimal balance between localization and mobility can maximize their chances to happen \cite{rivas20, buarque22}.
Here, this balance is effectively imposed by the correlation level $A$. To show there is indeed a 
coordinated dynamics taking place, let us track how many sites, for a given disordered sample, 
are mostly involved in the those extreme events. 
Defining $\pi (n)$ as the probability of a rogue occurring at site $n$, we compute the mean number of 
participating sites as $\Delta\overline{n}_{RW}=1/[\sum_n \pi(n)^2]$, which ranges from $1$ (fully biased) to $N$ (equally distributed). 
Figure \ref{fig5} displays this quantity against the correlation parameter $A$. 
Again, the non-monotonic behavior is clear and indicates that intermediate values of $A$ implies in more sites actively participating
in the generation of rogue waves. In turn, 
waves of higher amplitudes becomes more likely (cf. Figs. \ref{fig3} and \ref{fig4}).

Last, we complement the above discussion by taking a look back at the potential series given by Eq. (\ref{equation3}). Figure \ref{fig6}(a) 
shows how it typically sets up along the lattice. For $A=1$, the series features a white noise (almost uncorrelated) profile.
Then, at $A=25$ local fluctuations are drastically reduced. As we further
increase the effective correlation length, a rougher landscape is obtained, 
but still carrying a predominant harmonic component. 
To put all that together, let us compute 
the local disorder strength in terms of the local standard deviation $\sigma_{L_{0}} = \left(\sum_{k=1}^{M}\sigma_{k,L_{0}}\right)/M$ 
within a segment with $L_{0}$ sites, where 
\begin{equation}
\sigma_{k,L_{0}} = \sqrt{\sum_{n=(k-1)L_{0}+1}^{kL_{0}} \frac{\varepsilon_{n}^{2}}{L_{0}} - \left(\sum_{n=(k-1)L_{0}+1}^{kL_{0}}\frac{\varepsilon_{n}}{L_{0}}\right)^{2}} 
\label{eq:desvpfenergies}
\end{equation}
Results are shown in Fig. \ref{fig6}(b), where we are able 
to confirm once for all that intermediate values of $A$ makes for minimum local fluctuations, which is consistent
with the smooth potential landscape seen in Fig.\ref{fig6}(a).

\section{Concluding remarks}

We explored the intrinsic role played by correlated disorder on the emergence of rogue waves
in a simple quantum tight-binding model. The amplitude of the particle wave function was found to exhibit strong anomalous fluctuations
inherently unpredictable in time and space. An approach based on extreme-value statistics revealed
that such fluctuations follow a Gumbel distribution, belonging to the same class as those reported in 
a quantum walk featuring uncorrelated disorder \cite{buarque22}. 

We learned that 
intermediate values of the correlation parameter $A$, acting here as an effective correlation length, 
brings down the local disorder strength. This properly enhances the mobility of the wavefunction (by spanning eigenstates with larger localization lengths)
and so the number of sites on which
rogue-wave events take place, what in turn amplifies their characteristic amplitude (almost twice as large when compared 
to uncorrelated scenarios \cite{rivas20, bonatto20, buarque22}. 

While disorder \textit{must} be present to promote random fluctuations of the wavefunction, the underlying single-particle eigenstates must be wide enough in order to allow the linear superposition of the many components needed to produce localized waves in space and time. A general result we can establish is that rogue waves are expected to occur more often in the regime of weak disorder. 
Our findings are general to a class of 
1D discrete disordered systems 
and bring about another perspective on their dynamics 
as well as 
reveal fundamental aspects of the role of randomness in the generation of rogue waves.

%


\section{Acknowledgments}
This work was supported by CAPES, CNPq, and FAPEAL (Alagoas State agency).

\bibliography{references}

\end{document}